\documentclass[twocolumn,english,aps,prb, superscriptaddress,final]{revtex4}
\usepackage{color}
\usepackage{amstext}
\usepackage{graphicx}

\makeatletter

\providecommand{\tabularnewline}{\\}

\@ifundefined{textcolor}{}
{%
 \definecolor{BLACK}{gray}{0}
 \definecolor{WHITE}{gray}{1}
 \definecolor{RED}{rgb}{1,0,0}
 \definecolor{GREEN}{rgb}{0,1,0}
 \definecolor{BLUE}{rgb}{0,0,1}
 \definecolor{CYAN}{cmyk}{1,0,0,0}
 \definecolor{MAGENTA}{cmyk}{0,1,0,0}
 \definecolor{YELLOW}{cmyk}{0,0,1,0}
}

\usepackage{dcolumn}
\usepackage{bm}
\usepackage{times}
\usepackage{babel}

\makeatother

\usepackage{babel}
\begin{document}

\title{Electrostriction coefficient of ferroelectric materials from \emph{ab
initio} computation }

\author{Z. Jiang }

\affiliation{Electronic Materials Research Laboratory\textendash Key Laboratory
of the Ministry of Education and International Center for Dielectric
Research, Xi'an Jiaotong University, Xi'an 710049, China}

\affiliation{Physics Department and Institute for Nanoscience and Engineering,
University of Arkansas, Fayetteville, Arkansas 72701, USA }

\author{R. Zhang}

\affiliation{Department of Physics, State Key Laboratory of Photoelectric Technology
and Functional Materials (Cultivation Base), Northwest University,
Xi'an 710069, China}

\author{F. Li }

\affiliation{Electronic Materials Research Laboratory\textendash Key Laboratory
of the Ministry of Education and International Center for Dielectric
Research, Xi'an Jiaotong University, Xi'an 710049, China}

\author{L. Jin}

\affiliation{Electronic Materials Research Laboratory\textendash Key Laboratory
of the Ministry of Education and International Center for Dielectric
Research, Xi'an Jiaotong University, Xi'an 710049, China}

\author{N. Zhang}

\affiliation{Electronic Materials Research Laboratory\textendash Key Laboratory
of the Ministry of Education and International Center for Dielectric
Research, Xi'an Jiaotong University, Xi'an 710049, China}

\author{D. Wang}

\email{dawei.wang@mail.xjtu.edu.cn}

\selectlanguage{english}%

\affiliation{Electronic Materials Research Laboratory\textendash Key Laboratory
of the Ministry of Education and International Center for Dielectric
Research, Xi'an Jiaotong University, Xi'an 710049, China}

\author{C.-L. Jia}

\affiliation{Electronic Materials Research Laboratory\textendash Key Laboratory
of the Ministry of Education and International Center for Dielectric
Research, Xi'an Jiaotong University, Xi'an 710049, China}

\affiliation{\textsuperscript{}Peter Grünberg Institute and Ernst Ruska Center
for Microscopy and Spectroscopy with Electrons, Research Center Jülich,
D-52425 Jülich, Germany}
\begin{abstract}
Electrostriction is an important material property that characterizes
how strain changes with the development of polarization inside a material.
We show that \textit{ab initio }techniques developed in recent years
can be exploited to compute and understand electrostriction of ferroelectric
materials. Here, electrostriction coefficients of ferroelectric BaTiO$_{3}$,
PbTiO$_{3}$, as well as dielectric BaZrO$_{3}$, are obtained and
analyzed. Possible causes of the difference between experimental and
numerical results are discussed. We also identified that relative
displacements between certain ions at a given polarization could be
a good indicator of a material's electrostriction property.
\end{abstract}
\maketitle

\section{Introduction}

Electrostriction effect describes how a material deforms as a polarization
develops inside it. Being one of the nonlinear phenomena particularly
important in ferroelectric materials, it is the physical basis for
their application in actuators, sensors,\cite{Uchino1996,Uchino1981}
phase shifters, and sonar projectors.\cite{Park2002} The effectiveness
of electrostriction is often gauged by the electrostrictive coefficients
$Q_{ijkl}$ ($i=x,y,z$ represents Cartesian coordinate) that characterizes
how the strain of a material changes quadratically with its polarization,
i.e., $x_{ij}=Q_{ijkl}P_{k}P_{l}$,\cite{Newnham1997,Newnham2005}
where $x_{ij}$ represents strain tensors (in the following we will
use Voigt notation for strain tensor, i.e., $x_{i}$ represents strain
with $i=1\dots6$). Electrostriction is a four-rank polar tensor,
which can be observed in essentially all crystals regardless of their
symmetries.

Development of materials with large electrostriction is an important
task in material science research. For instance, it was suggested
that the large piezoelectric coefficient of the lead-free ceramic
system BZT-BCT\cite{Ren2009,LiFei2014} is originated from its large
electrostriction coefficients. It is obvious that accurately obtaining
electrostriction coefficient is a key step in such research. Experimentally,
there are several ways to obtain electrostriction coefficient $Q$:\cite{Newnham2005}
(i) Direct approaches, such as measuring the change in strain as a
function of polarization, or (ii) Indirect approaches, such as measuring
the change of permittivity (via change of capacitance) as a function
of mechanical stress. Computationally, the electrostriction coefficient
can be obtained by applying an electric field to the target material,
and compute the change of polarization and strain at each step as
the electric field increases. 

However, for ferroelectric materials, \textit{ab initio} computation
of electrostriction coefficients is not as straightforward as one
may expect even after the modern theory of polarization was established.\cite{Vanderbilt1993,Raffaele Resta_1994}
There are two main obstacles: (i) The application of an electric field
in a periodic structure needs to be treated carefully;\cite{Sai2002,Stengel2009}
(ii) Electrostriction is often obtained by assuming the target material
is in high symmetry phases, either because it is above the phase transition
temperature, $T_{C}$ (above which many perovskite ferroelectrics
are in the paraelectric phase), or because polycrystals or ceramics
are used, in which the overall phase is isosymmetric. Such a situation
causes the second difficulty in \textit{ab initio} calculation because
the mimicked high symmetry phase is \emph{unstable} for ferroelectric
materials (e.g., the paraelectric cubic phase for BaTiO$_{3}$ is
not the ground state). Such an unstable system can hardly be used
in \emph{ab initio} computation as it quickly jumps to other more
stable phases (e.g., the tetragonal phase in some perovskites) in
structural relaxations, forbidding us to set the system at a given
polarization and track how the strain changes.

Recently, these two obstacles have been overcome due to the progress
in \textit{ab initio} computation methods, which makes the calculation
of electrostriction coefficients easier. The first breakthrough came
when Souza \textit{et al.} and Umari \emph{et al}.\cite{Vanderbilt2002,Umari2002}
implemented the code for applying electric field in periodic structures.
With this advancement, it is possible to compute in general how a
material responds to applied electric field. As a matter of fact,
such method of fixing the electric field $\mathbf{E}$ has been applied
widely in the study of dielectric, piezoelectric, and ferroelectric
behavior of materials.\cite{Fu2003,Veithen2004,Antons2005,Stengel2007}
To address the second problem of the unstable high-symmetry phase,
Dieguez \textit{et al}.\cite{Dieguez2006} developed a method that
is able to impose a constant polarization, $\mathbf{P}$ (fixed-\textit{P}
method). In this way, one can increase the polarization of the system
gradually and track the change of strain, which paves the way for
numerically obtaining electrostriction coefficient. However, further
developments show that a better way is to impose the constraint of
a constant electric displacement, $\mathbf{D}$, (fixed-\emph{D} method).\cite{Stengel2009}
Among other reasons, this approach addresses a few subtle points that
are not fully appreciated in previous studies. For instance, (i) The
fixed-\emph{D} method corresponds to imposing an open-circuit electrical
boundary conditions, which is especially useful for studying ferroelectric
capacitors,\cite{Stengel_1_2009} superlattices\cite{Wu2008,Wu2011,Stengel2012}
and metal-oxide interfaces;\cite{stengePRBl2007,stengel2011,Cancellieri2011}
(ii) This approach made it possible to mimic the unstable cubic phase
of ferroelectric materials, and extract useful parameters (e.g., electrostriction
coefficient). More importantly, with the implementation of such a
method in the ABINIT software package,\cite{Hong2011} it enables
more researchers to obtain electrostriction coefficients via \textit{ab
initio} computation.

Here we use the fixed-\emph{D} method to study electrostriction in
prototypical ferroelectric materials BaTiO$_{3}$(BTO) and PbTiO$_{3}$(PTO).
As a comparison, we also present results from BaZrO$_{3}$(BZO), which
has the cubic phase as its ground state. Our numerical results indeed
shows that this approach can obtain electrostriction coefficients
consistent with experimental results, therefore providing a valuable
predictive power in discovering materials with better electrostrictive
properties, and an insight to the factors that determine electrostriction
coefficients.

\section{Method}

The \emph{ab initio} calculations were performed using the open-source
ABINIT software package.\cite{Gonze2002} Our calculations were performed
within density-functional theory in the local-density approximation
(LDA)\cite{Perdew1992} and the projector-augmented-wave (PAW) method\cite{Blochl1994}
is used along with pseudopotentials implemented in the GBRV package.\cite{Garrity2014}
Ba \textit{5s 5p 5d 6s}, Ti \textit{3s 3p 4s 3d}, Pb \textit{5d 6s
6p}, Zr\textit{ 4s 4p 4d 5s} and O \textit{2s 2p} orbitals were treated
as valence orbitals. A plane wave basis with kinetic energy cutoff
of 40 Hartree was used to ensure the convergence in all the calculations.
$k$-point sampling of $6\times6\times6$ Monkhorst-Pack grid\cite{Monkhorst1976}
was used for all the 5-atom unit cell. The atomic coordinates of the
five-atom unit cell were relaxed until all atomic-force components
were smaller than $10^{-5}$ Hatree/Bohr, and the cell size-and-shape
was varied until all stress components were below $10^{-7}$ Hatree/Bohr$^{3}$.

\section{Results and discussion}

The fixed-\emph{D} calculation implemented by Stengel \textit{et al}.\cite{Stengel2009,Hong2011}
is exploited in this work, where we gradually increase the reduced
electric displacement field $d$ along the $\left[001\right]$ direction.
This method enables the relaxation of both atom positions and cell
size-and-shape of a given crystal at a given $d$, the flux of the
electric displacement field, defined as $d$ = $\mathbf{a}_{1}\times\mathbf{a}_{2}D/(4\pi)$,\cite{Stengel2009}
where $\mathbf{a}_{1}$ and $\mathbf{a}_{2}$ are the in-plane lattice
vectors, and $\mathbf{D}$ is the electric displacement field. At
each step of the calculation process, the internal energy of the system,
$U$, the polarization of the system, as well as the relaxed cell
(determined by $\mathbf{a}_{1}$, $\mathbf{a}_{2}$, and $\mathbf{a}_{3}$)
are computed. After the computation, we have the relation between
strain and polarization, which can be used to compute electrostriction
coefficients. Note that for the calculation of strain, we use the
initial relaxed cubic phase $a_{0}$ as the reference. The system
is initially set to be cubic ($Pm\bar{3}m$), but with the application
of the fixed-displacement-field $\mathbf{D}$ along the $\left[001\right]$
direction, it becomes tetragonal ($P4mm$). The electric field, polarization,
internal energy, ionic positions and lattice parameters are all varied
at a given value of $d$. To obtain electrostriction coefficients
of each system (BTO, PTO and BZO), a series of computations are performed
at given electric displacement, $d$, along the $z$ direction. Internal
energy, polarization and strain are then obtained in each process,
and related to the constrained variable, $d$. We therefore obtain
the out-of-plane strain, $x_{3}$, and the in-plane strain $x_{1}$
as functions of $d$, which are then fitted to calculate electrostriction
coeffcients. Moreover, in this process we also track ion displacements
versus $d$, which provides some insight to the electrostriction effects.
We note that to define response properties it is important to specify
the boundary condition or constraint used in their definition.\cite{Wu2005,Wang2010}
Here our calculation of electrostriction coeffcients is under the
constraint of zero stress.

\subsection{Ferroelectric materials BaTiO$_{3}$ and PbTiO$_{3}$}

\begin{figure}[h]
\noindent \centering{}\includegraphics[width=8cm]{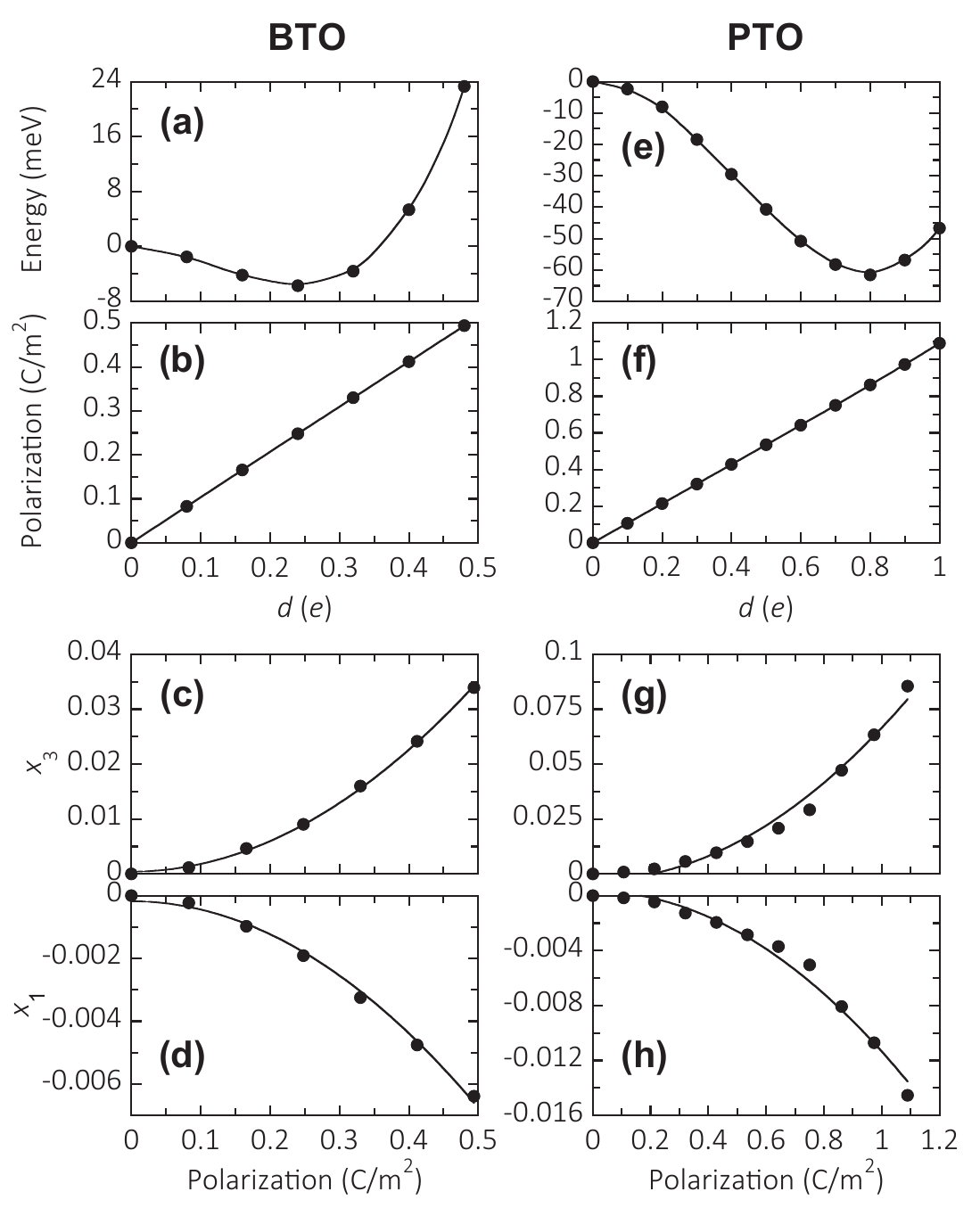}\caption{Internal energy (a), polarization (b), out-of-plane strain (c), and
in-plane strain (d) of BTO, as a function of $d$, the reduced electric
displacement {[}(a), (b){]} and polarization {[}(c), (d){]}. Internal
energy (e), polarization (f), out-of-plane strain (g), and in-plane
strain (h) of PTO, as a function of $d$ {[}(e), (f){]} and polarization
{[}(g), (h){]}. \label{fig:BTO-PTO} }
\end{figure}
We first show the results of internal energy \textit{$U$}, strain
in the $\left[001\right]$ ($x_{3}$) and $\left[100\right]$ direction
($x_{1}$) versus $d$, which lies in the $\left[001\right]$ direction.
For BTO, we start from the relaxed cubic structure (the lattice constant
$a_{c_{0}}=3.945\thinspace\textrm{\AA}$), and gradually increase
$d$ in steps of $0.08\thinspace e$ until it reaches 0.48 $e$ ($-e$
is the electron charge). For PTO, we increase $d$ in steps of $0.1\thinspace e$
until it reaches $1.0\thinspace e$. The structure was fully relaxed
with respect to both ionic positions and lattice parameters at each
$d$ value for both BTO and PTO. Strains are calculated using $x=(a_{c}-a_{c_{0}})/a_{c_{0}}$,
where $a_{c}$ is relaxed lattice constant at a given $d$. Numerical
results for BTO are shown in Fig. \ref{fig:BTO-PTO}. In Fig. \ref{fig:BTO-PTO}(a),
we plot the internal energy as a function of $d$. Clearly, there
is an energy minimum at $d\simeq0.24\thinspace e$, indicating that
the cubic BTO is unstable and such instability will introduce a ground
state with nonzero $d$. In Fig. \ref{fig:BTO-PTO}(b), we plot the
polarization versus $d$, and find that the polarization changes linearly
with $d$. The reason is that in the region of interest, it is shown
that the $\varepsilon_{0}E$ ($\varepsilon_{0}$ is the vacuum permittivity)
is much smaller than $P$. Therefore $P$ dominates the relation $D=\varepsilon_{0}E+P$,
ensuring this linear relation.\cite{Hong2011}\textcolor{red}{{} }Figures
\ref{fig:BTO-PTO}(c) and \ref{fig:BTO-PTO}(d) show that the strain
versus polarization curves are quadratic, which implies that electrostriction,
not piezoelectricity, dominates the mechanical response to polarization.
To quantify the electrostriction effect, we employ the expression
$x_{ij}=Q_{ijkl}P_{k}P_{l}$ to fit the curves in Figs. \ref{fig:BTO-PTO}(c)
and \ref{fig:BTO-PTO}(d).\cite{Newnham2005} More precisely, we use
the equation $x=a_{0}+a_{2}P^{2}$ for fitting. It is found that $Q_{11}=0.137$
m$^{4}$/C$^{2}$, which is quite close to the experimental value
of $Q_{11}=0.11$ m$^{4}$/C$^{2}$, and $Q_{12}=-0.026$ m$^{4}$/C$^{2}$,
which is comparable to the experimental value of $Q_{12}=-0.045$
m$^{4}$/C$^{2}$.\cite{Li2014,const-volume} The piezoelectric voltage
coefficients $g_{33}$ and $g_{31}$ can be obtained from the relation
$g_{ijk}=2Q_{ijkl}P_{s}$.\cite{Newnham1997} For BTO, we obtained
$g_{33}=0.071$ V m/N and $g_{31}=-0.014$ V m/N, which are comparable
to the experimental values of $g_{33}=0.058$ V m/N and $g_{31}=-0.023$
V m/N.\cite{Jaffe1958} This comparison supports the idea that electrostriction
may be used to estimate the electromechanical properties (e.g., piezoelectric
coefficients) of ferroelectric materials with spontaneous polarization. 

Numerical results for PTO is also shown in Fig. \ref{fig:BTO-PTO},
where we plot various properties of PTO under different $d$. Figure
\ref{fig:BTO-PTO}(e) shows that PTO also has an energy minimum, which
agrees well with previous studies\cite{Stengel2009,Hong2011} in terms
of the double-well potential and the position of the energy minimum.
Figure \ref{fig:BTO-PTO}(e) indicates that the depth of energy well
of PTO ($\sim60$ meV) is much deeper than that of BTO ($\sim5$ meV)
and the energy minimum corresponds to a larger polarization ($P=0.86$
C/m$^{2}$ versus $P=0.25$ C/m$^{2}$ in BTO.). In Figs. \ref{fig:BTO-PTO}(g)
and \ref{fig:BTO-PTO}(h), we show the out-of-plane and in-plane strain
versus polarization, and found $Q_{11}=0.065$ m$^{4}$/C$^{2}$ and
$Q_{12}=-0.012$ m$^{4}$/C$^{2}$ by fitting the two quadratic curves.
Comparable experimental values of $Q_{11}=0.09$\ m$^{4}$/C$^{2}$
and $Q_{12}=-0.03$\,\,m$^{4}$/C$^{2}$ have been reported.\cite{Li2014}
We note that the fitting range and whether including higher order
terms can potentially affect the final results. For instance, if we
include the $P^{4}$ term to fit Figs. 1(g) and 1(h), we find $Q_{11}=0.046$
m$^{4}$/C$^{2}$ and $Q_{12}=-0.009$ m$^{4}$/C$^{2}$.

\subsection{Dielectric material BaZrO$_{3}$}

\begin{figure}[h]
\noindent \centering{}\includegraphics[width=8cm]{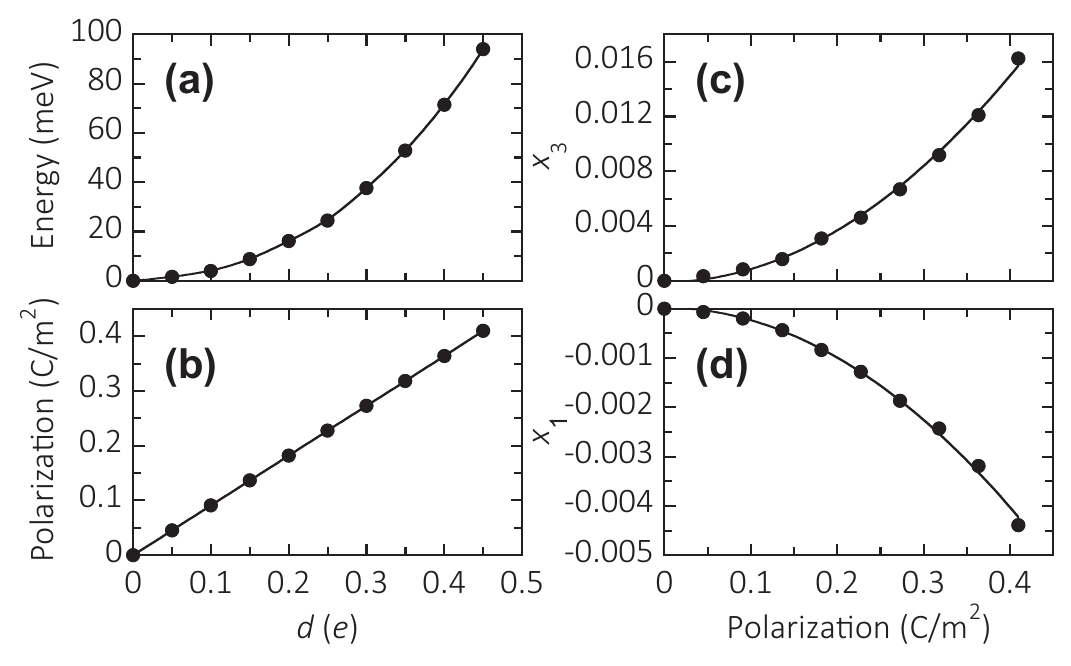}\caption{Internal energy (a), polarization (b), out-of-plane strain (c), and
in-plane strain (d) of BZO, as a function of $d$ {[}(a), (b){]} and
polarization {[}(c), (d){]}.\label{fig:BZO}}
\end{figure}
Having examined ferroelectric materials, we now move to BZO, which
is a dielectric material with perovskite structure that has larger
lattice constant, smaller thermal expansion, and smaller dielectric
permittivity than BaTiO$_{3}$.\cite{Dobal2001,Azad2002,Akbarzadeh2005}
Experimental and theoretical results indicate that BZO remains paraelectric
at low temperatures,\cite{Akbarzadeh2005} which means it is not piezoelectric,
but can have electrostriction effect. Figure \ref{fig:BZO} shows
the internal energy $U$ and polarization versus $d$, and strains
versus polarization. Figure \ref{fig:BZO}(a) shows that, unlike BTO
and PTO, the internal energy $U$ always increases with $d$. This
is consistent with the the fact that cubic BZO has no ferroelectric
instability. The electrostriction coefficients we obtained are $Q_{11}=0.0944$
m$^{4}$/C$^{2}$ and $Q_{12}=-0.0254$ m$^{4}$/C$^{2}$. Since direct
experimental value on $Q_{11}$ and $Q_{12}$ are not available, we
also use $Q_{h}=Q_{11}+2Q_{12}$ to calculate the hydrostatic electrostrictive
coefficients and find $Q_{h}=0.0436$ m$^{4}$/C$^{2}$. As a comparison,
the experimental value of $Q_{h}$ is found to be $0.023$ m$^{4}$/C$^{2}$
in Ref. \cite{Uchino1980} where the technique of measuring capacitance
change under hydrostatic pressure is used. The difference between
experimental $Q_{h}$ and the value we find here may be related to
the facts that (i) We did not use the constant volume constraint,\cite{const-volume}
and/or (ii) There are anti-phase motions of oxygen octahedra found
in BZO.\cite{Akbarzadeh2005,Lebedev2015} It is interesting to note
that the numerical $Q_{11}$ of BZO obtained here is larger than PTO
(but smaller than BTO). However, it is worth noting that since BZO
is not ferroelectric, it may be difficult to convert the large electrostriction
effect into piezoelectricity, unlike BTO or PTO in which spontaneous
ferroelectricity automatically break the centrosymmetry without external
electric field.

Since BZO does not have a ferroelectric instability, it is also possible
to obtain its electrostriction coefficient by directly applying an
electric field. Here, we apply an electric field, $E$, along the
$\left[001\right]$ direction. We applied electric field up to $E=1.14\times10^{9}$
V/m. At every step of the applied electric field, the structure of
BZO, including its lattice constant, atomic positions, and cell shape,
are optimized, similar to what is done in the fixed-\emph{D} approach.
Once the optimized structure of BZO at a given $E$ is obtained, its
polarization is calculated using the Berry phase approach.\cite{Vanderbilt1993}
In this way, the electrostriction coefficients are found to be $Q_{11}=0.0946$
m$^{4}$/C$^{2}$ and $Q_{12}=-0.0265$ m$^{4}$/C$^{2}$, which agree
well with the results obtained via the fixed-\emph{D} method.

\subsection{Ion displacement}

\begin{figure}[h]
\noindent \centering{}\includegraphics[width=8cm]{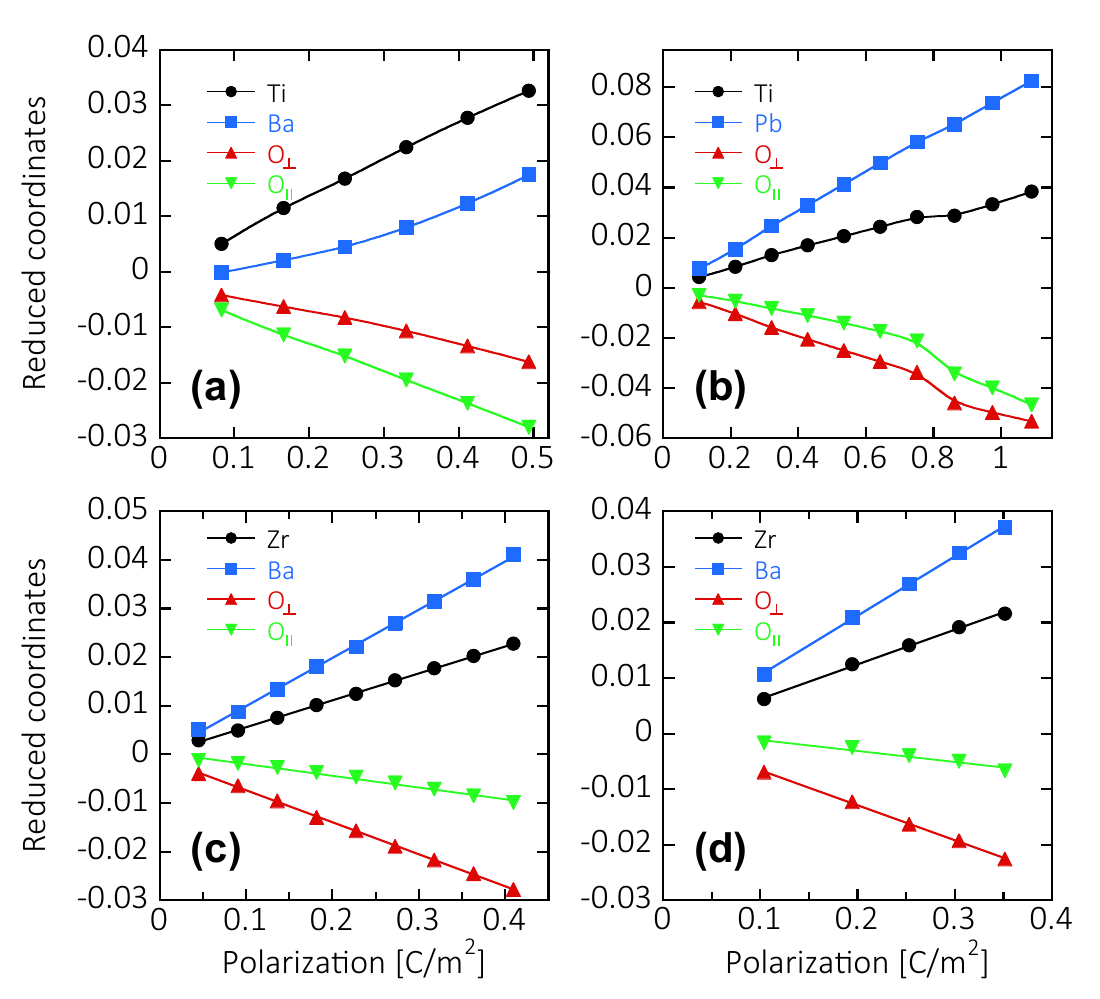}\caption{The out-of-plane displacements of each ion in BTO, PTO and BZO from
zero-field positions change with polarization obtained with the fixed-\emph{D}
method are shown in panels (a), (b), and (c), respectively. For BZO,
the method of directly applying an electric field is also shown {[}Panel
(d){]}.\label{fig:ion-shifts} }
\end{figure}
We also obtained the displacement of each ions versus polarization
in Fig. \ref{fig:ion-shifts}(a), where we show the reduced coordinates
of BTO versus the polarization. A prominent feature of this figure
is that all the atoms shift linearly with polarization. As expected,
Ba and Ti ions (being positively charged) move along the direction
of the reduced electric displacement field $d$, while O$_{\perp}$
and O$_{\parallel}$ ions (note the Ti-O$_{\parallel}$ and Ti-O$_{\perp}$
bonds are parallel and perpendicular to the $z$ direction, respectively),
which are negatively charged, move in the opposite direction. Together,
they induce a polarization along the $z$ direction. We further note
that the displacements of Ti, Ba, O$_{\perp}$, O$_{\parallel}$ obtained
here with the fixed-\emph{D} method are compatible with that obtained
from force-constant matrix, where the eigenvector is given by $\left(\xi_{\textrm{Ba}},\xi_{\textrm{Ti}},\xi_{\textrm{O}_{\perp}},\xi_{\textrm{O}_{\parallel}}\right)=\left(0.20,0.76,-0.21,-0.53\right)$.\cite{Smith1994}
Here we find from the slope of the lines shown in Fig. \ref{fig:ion-shifts}(a)
that the eigenvector is $\left(0.20,0.66,-0.32,-0.57\right)$. The
closeness of these two results and the linear correlation shown in
Fig. \ref{fig:ion-shifts}(a) further validate the accuracy of the
effective Hamiltonian approach in the study of BTO.\cite{Smith1994,Zhong1995}
It is worth noting that the eigenvector shown in Ref. \cite{Smith1994}
is obtained by diagonalizing the force-constant matrix. With a series
of similar fixed-$D$ calculations, we believe the static field-induced
ion displacements agree better with the force-constant matrix eigenvectors
than the dynamical matrix eigenvector. Figure \ref{fig:ion-shifts}(b)
shows the displacement of each ions in PTO as a function of the polarization.
Similar to BTO, the displacements of the ions are linearly dependent
on $P$, and consistent with the eigenvector obtained in frozen phonon
calculation \cite{Smith1994} and experiments.\cite{Zhang2011} However,
one can see that in PTO, the A-site ion Pb moves in a larger distance
than the B-site ion, Ti, which is opposite to what happens in BTO.
Since the displacement of Pb ion is about twice as large as that of
Ti, this result also indicates that in PTO Pb has larger contribution
to polarization than Ti. Since Pb is outside oxygen octahedrons, it
likely has more room to move without distorting the unit cell along
$z$ , unlike Ti ions. In other words, to achieve the same polarization,
Ti in PTO has a smaller displacement than BTO, and thus a smaller
deformation, which results in a smaller $Q_{11}$. Such a difference
may explain why $Q_{11}$ of PTO is smaller than that of BTO. We thus
suggest that B-site driven ferroelectrics may in general have better
electrostriction than A-site driven ones. Figure \ref{fig:ion-shifts}(b)
also shows an anomalous change around $P=0.8\thinspace\textrm{C}/\textrm{m}^{2}$.
The inverse capacitance of PTO was also found to have such a change,
which is discussed in Ref. \cite{Stengel2009}.

\begin{figure}[h]
\noindent \centering{}\includegraphics[width=8cm]{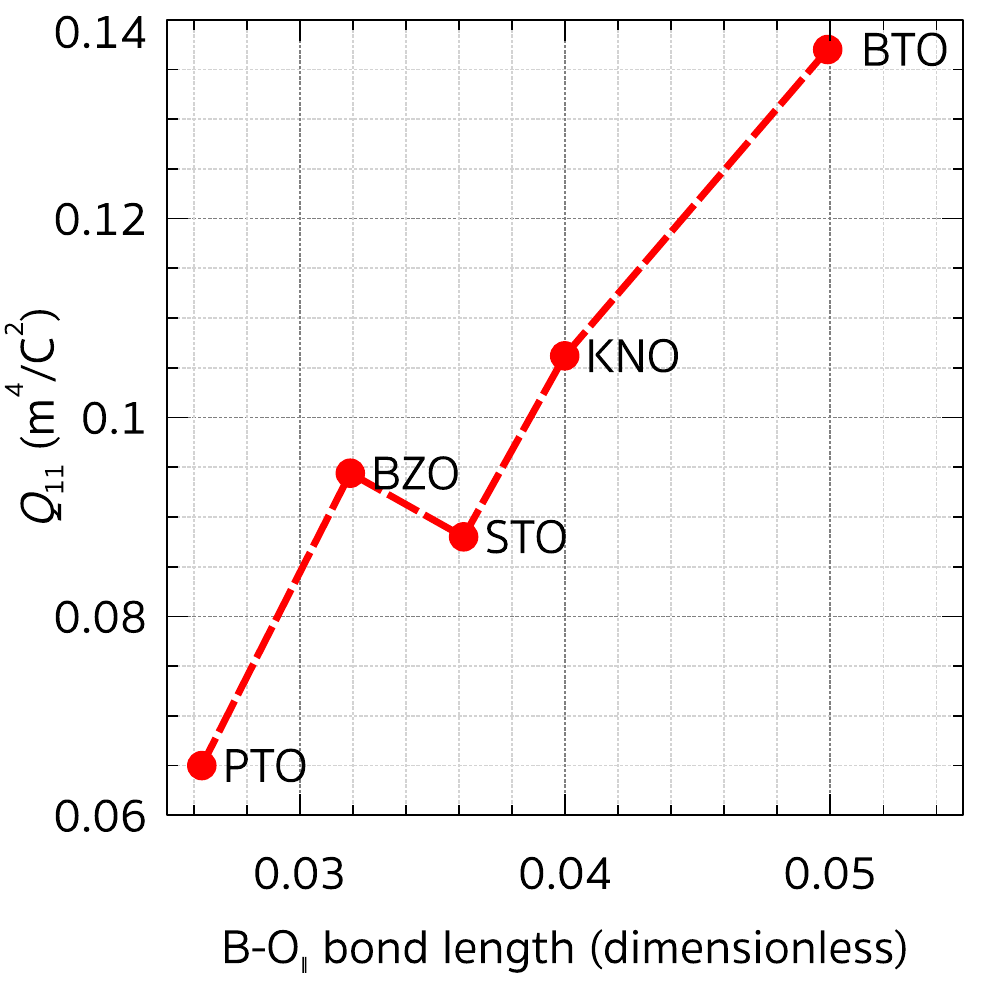}\caption{The \emph{B}-O$_{\parallel}$ bond length (in reduced coordinates)
at $P=0.4\thinspace\textrm{C}/\textrm{m}^{2}$ and $Q_{11}$ of BTO,
KNO (KNbO$_{3}$), STO (SrTiO$_{3}$), BZO and PTO.\label{fig:bond-distance}}
\end{figure}
The ion displacements in BZO versus $d$ also has a linear behavior
as shown in Fig. \ref{fig:ion-shifts}(c) (fixed-D) and Fig. \ref{fig:ion-shifts}(d)
(fixed-\emph{E}), where Ba and Zr ions move along $d$ or $E$, while
O ions move in the opposite direction. It is worth noting that, in
BZO, the A-site atom, Ba, shifts more than the B-site atom, Zr, which
is opposite to what happens in BTO, but similar to PTO. Comparison
of Fig. \ref{fig:ion-shifts}(c) and Fig. \ref{fig:ion-shifts}(b)
to Fig. \ref{fig:ion-shifts}(a) also reveals another important difference
between BTO and BZO/PTO, i.e., $\left|\xi_{O_{\parallel}}\right|>\left|\xi_{O_{\perp}}\right|$
in BTO while the opposite is true in BZO/PTO. The larger $\xi_{O_{\perp}}$
in BZO/PTO is likely due to the larger displacement of the A-site
atom (Ba in BZO and Pb in PTO) than that of the B-site atom ( Zr in
BZO and Ti in PTO) as the displacement A-site ion may cause neighboring
oxygen octahedrons to tilt. The importance of such difference on electrostriction
becomes apparent when the bond length of the B-site atom and O$_{\parallel}$
along with $Q_{11}$ is summarized in Fig. \ref{fig:bond-distance}
for five different perovskites (Table \ref{tab:parameters} summarizes
the numerical results of $Q_{11}$, $Q_{12}$, $g_{33}$, $g_{31}$,
$C_{11}$, $C_{12}$, $C_{44}$ and spontaneous polarization $P$,
the results of elastic constants and spontaneous polarization are
in general agree with existing computational and experimental results.\cite{Smith1994,Wang2010,Nishimatsu2010,Berlincourt1958,Liu2008,Terki2005,Li1996,Yamanaka2003,Kalinichev1993,Wieder1955,Fechner2008,Kleemann1984,Gavrilyachenko1970,Wang1997}),
which indicates that $Q_{11}$ has positive correlation with the \emph{B}-O$_{\parallel}$
bond length. In other words, longer \emph{B}-O$_{\parallel}$ means
larger theoretical electrostriction coefficients when other things
(e.g. polarization) being equal. This relation can be understood by
realizing the fact that the \emph{B}-O$_{\parallel}$ bond length
characterizes and, to some extent, determines the unit cell deformation
along $z$, which in turn decides the magnitude of $Q_{11}$. In practice,
such an indicator can potentially facilitate the investigation of
electrostriction by paying more attention to the \emph{B}-O$_{\parallel}$
bond length.
\begin{table}[h]
\noindent \begin{centering}
\begin{tabular}{cccccc}
\hline 
 & BaTiO$_{3}$ & KNbO$_{3}$ & SrTiO$_{3}$ & BaZrO$_{3}$ & PbTiO$_{3}$\tabularnewline
\hline 
$Q_{11}$ (m$^{4}$/C$^{2}$) & 0.137 & 0.106 & 0.088 & 0.094 & 0.065\tabularnewline
\hline 
$Q_{12}$ (m$^{4}$/C$^{2}$) & -0.026 & -0.028 & -0.020 & -0.025 & -0.012\tabularnewline
\hline 
$g_{33}$ (V m/N) & 0.071  & 0.066 & - & - & 0.108\tabularnewline
\hline 
$g_{31}$ ( V m/N) & -0.014 & -0.017 & - & - & -0.020\tabularnewline
\hline 
$C_{11}$ (GPa) & 322.7 & 422.6 & 355.2 & 332.4 & 324.5\tabularnewline
\hline 
$C_{12}$ (GPa) & 110.3 & 73.6 & 101.6 & 79.9 & 124.0\tabularnewline
\hline 
$C_{44}$ (GPa) & 126.6 & 94.1 & 113.9 & 86.6 & 101.1\tabularnewline
\hline 
$P_{s}$ (C/m$^{2}$) & 0.26 & 0.31 & 0.00 & 0.00 & 0.83\tabularnewline
\hline 
B-O$_{\parallel}$ & 0.050 & 0.040 & 0.036 & 0.032 & 0.026\tabularnewline
\hline 
\end{tabular}
\par\end{centering}

\caption{The electrostriction coefficients $Q_{11}$, $Q_{12}$, piezoelectric
voltage coefficients $g_{33}$ and $g_{31}$, cubic phase elastic
constants $C_{11},$ $C_{12}$, $C_{44}$, spontaneous polarization
$P_{s}$ , and the B-O$_{\parallel}$ bond length (in reduced coordinates)
at $P=0.4$ C/m$^{2}$ for BTO, KNO, STO, BZO and PTO.\label{tab:parameters} }
\end{table}

\subsection{Discussion \label{sub:Discussion}}

With the aid of \textit{ab initio} computation in general and the
fixed-\emph{D} method in particular, we now have an additional perspective
to understand electrostriction. As we know, electrostriction represents
a relation between displacements of atoms (that determines the polarization)
and structural parameters (that determines strain). It is often taken
for granted that the polarization is caused by an external electric
field. However, it has been shown that the relation between $\mathbf{P}$
and $\mathbf{E}$ is rather complex \cite{Stengel2009,Hong2011} for
ferroelectric materials. For instance, to constrain the system to
a given $\mathbf{P}$, $\mathbf{E}$ need to be negative (i.e., opposite
in the direction of $\mathbf{P}$) to stabilize the system at a given
configuration. The reason is not hard to understand \textendash{}
it is the internal electric field arising from electric dipoles that
drives the ferroelectric instability, which in turn results in the
spontaneous polarization.\cite{Vanderbilt1993,Zhong1995} For ferroelectric
materials, it is also found from \textit{ab initio} computation that
for $\mathbf{P}$ values of our interest, the required electric field
is much smaller than $P$, i.e., $\varepsilon_{0}E\ll P$,\cite{Hong2011}
which explains why the relation $x_{ij}=Q_{ijkl}P_{k}P_{l}$ is adopted
more than $x_{ij}=M_{ijkl}E_{k}E_{l}$ to characterize electrostriction
in ferroelectric materials.

As we have shown, the fixed-\emph{D} approach is effective to obtain
electrostriction coefficients via \emph{ab initio} computations. This
computational approach is obviously useful for discovering materials
or structures with desired electrostrictive properties. In order to
employ this method properly, it is important to understand potential
causes of difference between experimental work and \emph{ab initio}
calculation. First, in many experiments on electrostriction, polycrystals
or ceramics are used, which has two implications: (i) The effect of
domains, domain walls, and/or grain boundaries in real materials cannot
be reflected in \textit{ab initio} calculation with a rather small
number of unit cells; and (ii) The sample may not be in its high symmetry
phase due to mechanical stress or low temperature (lower than $T_{C}$)
\textendash{} for such a situation, a direct comparison is not ideal.
More sophisticated \emph{ab initio }computation of other phases (e.g.
the tetragonal phase) and some statistics on the results may be mandatory.
Second, in comparison with experimental values, one needs to be aware
of some electric boundary conditions between different \emph{ab initio}
computations. For instance, the fixed-$E$ computation corresponds
to a closed circuit electric boundary condition with a biased voltage
and assumes that $E$ is continuous through the system, while the
fixed-\emph{D} method corresponds to an open circuit electric boundary
condition and continuity of $\mathbf{D}$.\cite{Stengel2009,Vanderbilt2002,Umari2002}
This subtlety warrants careful consideration before adopting one or
the other approach. For complex non-uniform structures, we suggest
to use the fixed-\emph{D} approach, which is particularly important
for layered structures, such as superlattices or metal/insulator interfaces,
as pointed in Ref. \cite{Stengel2009}. This practice should be considered
even for non-ferroelectric materials where applying electric field
is not a problem. Third, more complex structures (e.g. the anti-phase
oxygen octahedron tiltings in BZO) could affect the accuracy of \textit{ab
initio} results. However, we note that introducing such complex crystal
structures will substantially slow down the calculation with the fixed-\emph{D}
approach. Finally, \textit{ab initio} computations assume zero temperature,
therefore the optimized lattice constant may not corresponds to the
experimental values at finite temperatures. Experimentally verified
values (e.g., lattice constant) may be needed to reduce differences
in numerical values.

\section{Conclusion}

In summary, we have shown that the fixed-\emph{D} approach can be
applied to obtain electrostriction coefficients of ferroelectric materials.
By investigating three materials (BTO, PTO, and BZO), we find the
fixed-\emph{D} approach provide a powerful method to compute and understand
electrostrictive effects of various materials, and is potentially
useful for future predictions of materials or designs. With the aid
of this approach, we have also shown that, given a perovskite material,
the bond length of the B-site atom and the O$_{\parallel}$ ion is
an important indicator of the magnitude of its electrostriction, which
could be a useful criterion in screening a large number of perovskites
to find those with good electrostrictive effects. In addition, possible
difference between electrostriction coefficients obtained theoretically
and experimentally are discussed in detail. We thus hope that the
approach we demonstrat here will be widely used in discovering and
exploring materials with better electrostrictive effects.
\begin{acknowledgments}
This work is financially supported by the National Natural Science
Foundation of China (NSFC), Grant No. 51390472, 11574246, and National
Basic Research Program of China, Grant No. 2015CB654903. F.L. acknowledges
NSFC Grant No. 51572214. Z.J. also acknowledges support from China
Scholarship Council (CSC No. 201506280055) and the 111 Project (B14040).
We thank Shanghai Supercomputer Center for providing resources for
\textit{ab initio} computations. \end{acknowledgments}


\begin{thebibliography}{10}
\bibitem{const-volume}We note that if constant volume is assumed
in the process of measuring electrostriction coefficients, it can
be shown that, for the tetragonal symmetry, $Q_{12}\simeq-\frac{Q_{11}}{2}(1-\frac{3Q_{11}}{4})$.
We did not use this constraint in structural relaxations since such
an assumption is not suitable for all situations. For instance, for
thin films grown on a substrate, the volume can indeed change with
$d$ applied in the out-of-plane direction.
\end{thebibliography}
\end{document}